\documentclass[aip,jcp,reprint]{revtex4-1}

\usepackage{graphicx}

\usepackage{amsmath}

\begin{document}

\title{Raman and nuclear magnetic resonance investigation of alkali metal vapor interaction with alkene-based anti-relaxation coating}

\author{O.\ Yu.\ Tretiak}
\email[]{otretiak@genphys.ru}
\affiliation{Physical faculty, St. Petersburg State University, 7/9 Universitetskaya nab., St. Petersburg, 199034, Russia}

\author{J.\ W.\ Blanchard}
\affiliation{Department of Chemistry, University of California at Berkeley, CA, 94720, USA}
\affiliation{Materials Sciences Division, Lawrence Berkeley National Laboratory, Berkeley, CA, 94720, USA}
\affiliation{Helmholtz Institute Mainz, Johannes Gutenberg University, 55099 Mainz, Germany} 

\author{D.\ Budker}
\affiliation{Helmholtz Institute Mainz, Johannes Gutenberg University, 55099 Mainz, Germany} 
\affiliation{Department of Physics, University of California at Berkeley, CA, 94720-7300, USA}
\affiliation{Nuclear Science Division, Lawrence Berkeley National Laboratory, Berkeley, CA, 94720, USA}

\author{P.\ K.\ Olshin}
\affiliation{Resource Center ``Optical and laser methods of material research", St. Petersburg State University, 7/9 Universitetskaya nab., St. Petersburg, 199034, Russia}

\author{S.\ N.\ Smirnov}
\affiliation{Center for Magnetic Resonance, St. Petersburg State University, 7/9 Universitetskaya nab., St. Petersburg, 199034, Russia}

\author{M.\ V.\ Balabas}
\affiliation{Physical faculty, St. Petersburg State University, 7/9 Universitetskaya nab., St. Petersburg, 199034, Russia}

\date{\today} 

\begin{abstract}
The use of anti-relaxation coatings in alkali vapor cells yields substantial performance improvements by reducing the probability of spin relaxation in wall collisions by several orders of magnitude. Some of the most effective anti-relaxation coating materials are alpha-olefins, which (as in the case of more traditional paraffin coatings) must undergo a curing period after cell manufacturing in order to achieve the desired behavior. Until now, however, it has been unclear what physicochemical processes occur during cell curing, and how they may affect relevant cell properties. We present the results of nondestructive Raman-spectroscopy and magnetic-resonance investigations of the influence of alkali metal vapor (Cs or K) on an alpha-olefin, 1-nonadecene coating the inner surface of a glass cell. It was found that during the curing process, the alkali metal catalyzes migration of the carbon-carbon double bond, yielding a mixture of cis- and trans-2-nonadecene.
\end{abstract}


\maketitle

\section{Introduction\label{intro}} 

Alkali-metal vapor cells are used in quantum-optics and metrology  experiments \cite{Krauter2013,Sherson2006,Julsgaard2001}. Among the key parameters in these experiments are alkali metal vapor density and ground state  atomic
polarization relaxation time \cite{Budker2007}. Anti-relaxation coatings of the inner surfaces of alkali-metal vapor cells are used to decrease the probability of spin relaxation in a wall collision by up to six orders of magnitude compared to the case of uncoated glass (see review by Balabas, Bouchiat, and Seltzer in the Atomic Magnetometry book that now appears as Ref. \cite{Jackson2013}).
 
At the same time, anti-relaxation coating can absorb metal atoms and decrease vapor density \cite{Bouchiat1966,Balabas1993,Balabas2012,Balabas2013}.  This process is most intense during the ``curing period'' just after the cell manufacturing. Before the cell is cured, the vapor density in the cell is suppressed, and the relaxation rates are high.  The mechanisms  of the curing process and their role in the final properties of a cell such as relaxation time and alkali-metal equilibrium vapor pressure are still unknown.

Traditionally, anti-relaxation coatings have been composed of a mixture of alkanes \cite{Robinson1958}, but the longest relaxation times achieved in alkali vapor cells (up to 77 s) have made use of alkene-based cell coatings \cite{PhysRevLett.105.070801}. It is not a-priori clear why an unsaturated hydrocarbon-based coating would show superior anti-relaxation properties.

  It remains difficult to predict the exact chemical structure of the final product of the interaction of the hydrocarbon with an alkali metal, because the  chemistry at the interface between the alkali metal vapor and the organic coating material has not been fully explored. Mechanisms for improvement of the quality of alkene anti-relaxation coatings involving chemical binding of cesium atoms and creation of metal-organic compounds were proposed in Ref.~\cite{Balabas2010}.

In the present work, we study alkali-vapor interaction with a particularly successful  \cite{PhysRevLett.105.070801} alkene coating material (1-nonadecene) by Raman and nuclear magnetic resonance (NMR) spectroscopy methods. These methods were chosen as  non-destructive methods, allowing for in-situ analysis.  The results of these studies allow us to identify the nature of the chemical modification occurring during curing.

\section{EXPERIMENTAL METHODS\label{METHODS}}

\begin{figure}
	\centering
		\includegraphics[width=0.9 \linewidth]{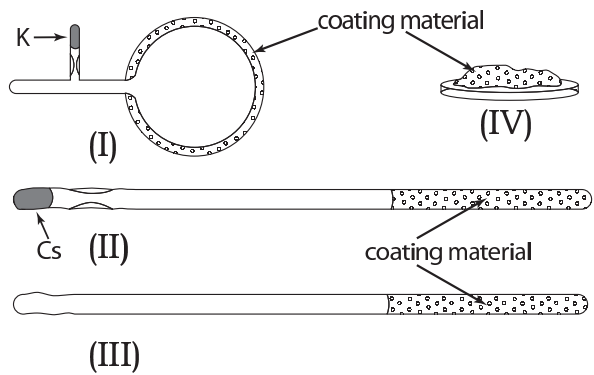}
	\caption{Experimental samples: I -- potassium vapor cell with anti-relaxation coated walls, II -- evacuated standard 5~mm NMR tube with coating material and Cs drop, III -- 
	an otherwise identical evacuated NMR tube with coating material, but without Cs drop,
	IV -- coating material on glass wafer.
	} \label{fig:tube}
\end{figure}

Raman spectroscopy is useful for investigating anti-relaxation coating through the glass, inside an alkali-metal vapor cell (Fig.~\ref{fig:tube}~I) because of the high transparency  of pyrex glass  for the pumping light. While infrared (IR) spectroscopy cannot be used because of IR opacity of the glass and X-ray photoelectron spectroscopy (XPS) requires disassembling vapor cell in vacuum for direct access to the coating.

All samples in Fig.~\ref{fig:tube} were investigated by Raman spectroscopy at the Saint-Petersburg State University Center for Optical and Laser Methods of Material Research [http://researchpark.spbu.ru/laser]  using an express-Raman spectrometer SENTERRA (Bruker). 
This apparatus allows for spatial profiling of spectra with micron resolution.
Fluorescence background in the Raman spectra was suppressed with the spectrometer software using wavelength sweeping of pumping diode laser.

Samples II and III (Fig.~\ref{fig:tube}) were prepared in standard 5mm Pyrex NMR tubes.
Both samples were evacuated and flame-sealed under vacuum.
A typical procedure of the coated cell preparation is described in \cite{Budker2013}.
The same fraction of alpha-olefin as in  \cite{Balabas2010} was used. The initial material was Alpha-Olefin fraction C~20-24 from Chevron Phillips (CAS Number 93924-10-8). A light fraction of the material was removed through vacuum distillation at $T=80~\rm^{\circ}C$. The remains were used as the coating material.

$^{1} \rm H$ and $^{13} \rm C$ NMR experiments were performed at the Saint-Petersburg State University Center for Magnetic Resonance [http://researchpark.spbu.ru/~cmr] with a Bruker~300DPX spectrometer. The instrumental settings were:  for $^{1} \rm H$ (300.13~MHz):  20 ~ppm spectral width, acquisition time 1.34~s, single scan; for $^{13} \rm C$ (75.74~MHz): 235 ~ppm spectral width, acquisition time 0.93~s, relaxation delay 6~s, the signal was averaged over 7500 transients. 
All spectra were recorded without sample spinning above the melting point of the alkenes at $38 \rm ^\circ C$.

\begin{figure}
	\centering
		\includegraphics[width=\linewidth]{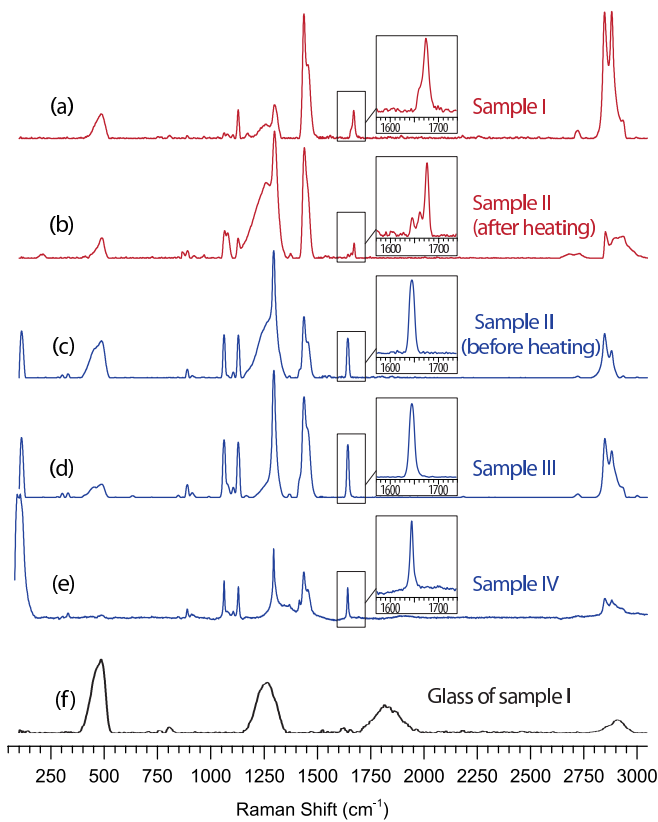}
	\caption{Raman spectra (recorded at $\rm 21~^\circ C$):  a -– potassium vapor cell with alkene anti-relaxation coated walls (Fig.~\ref{fig:tube}~I), b -- NMR tube with the coating material and Cs vapor (Fig.~\ref{fig:tube}~II) after heating at $80 \rm ^\circ C$ during one month, c -- NMR tube with the coating material and Cs vapor (Fig.~\ref{fig:tube}~II) immediately after the preparation, d -– control NMR tube with the coating material (Fig.~\ref{fig:tube}~III) after heating in the same conditions as with the Cs drop above, e -- alkene coating on a glass wafer in the air (Fig.~\ref{fig:tube}~IV), f -- glass wall of the cell (Fig.~\ref{fig:tube}~I). All spectra were normalized to the highest peak. The peaks characteristic of the $\rm C=C$ bond are labeled with rectangles and shown zoomed in the insets.
	}
	\label{fig:r-spectra}
\end{figure}

All Raman spectra shown in Fig.~\ref{fig:r-spectra} except the top one (a) were recorded  with the same instrumental settings: pump wavelength was 785~nm and 100~mW power with automatic fluorescence rejection. 
Fig.~\ref{fig:r-spectra}a alkene coating on glass was recorded at a wavelength of 785~nm and 25~mW power, fluorescence subtracted manually.  Comparison of spectra from different samples reveals the influence of alkali metal atoms and air on the coating material under consideration. 

\section{EXPERIMENTAL RESULTS: RAMAN \label{RAMAN}} 

The results of the Raman spectroscopy investigations are summarized in Fig.~\ref{fig:r-spectra} which shows the observed spectral lines in the range relevant to hydrocarbons  \cite{Socrates2004}. While the spectra for different coatings and conditions are generally similar, there are also notable differences.
As mentioned in  \cite{Balabas2010}, the  peak at 1643 $ \rm cm^{-1}$ 
corresponding to the C=C stretching mode, disappeared after a long-term interaction between potassium vapor and the coating. 
The peak characteristic of the terminal $\rm C=C$ bond is zoomed  for each of the spectra shown in Fig. 2. For the initial material (Fig.~\ref{fig:r-spectra}d, \ref{fig:r-spectra}e) there is single peak at $\rm 1643~cm^{-1}$ like in \cite{Balabas2010}. The same result was obtained for the tube with a Cs drop immediately after its preparation (Fig.~\ref{fig:r-spectra}c). However, upon exposure to alkali-metal vapor, we see a decrease in the $\rm 1643~cm^{-1}$ line accompanied by the appearance of stronger lines at $\rm 1659~cm^{-1}$  ad $\rm 1673~cm^{-1}$ (Fig.~\ref{fig:r-spectra}a and  Fig.~\ref{fig:r-spectra}b).  

\begin{figure}
	\centering
		\includegraphics[width=0.9 \linewidth]{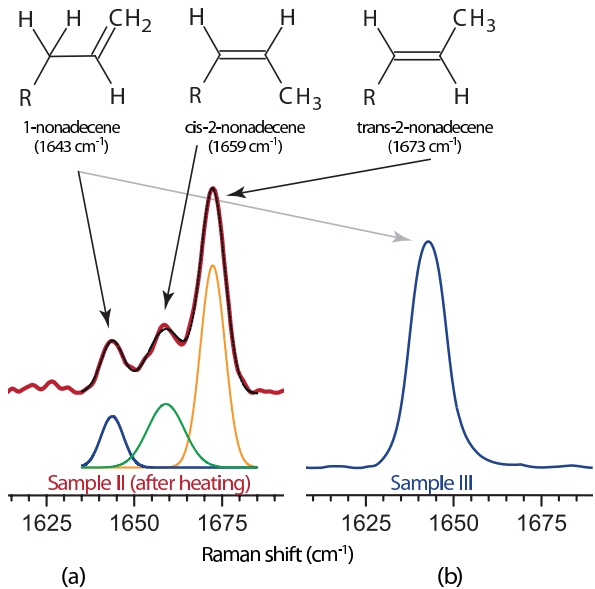}
	\caption{Gaussian fits of the $\rm 1643~cm^{-1}$,  $\rm 1659~cm^{-1}$ and $\rm 1673~cm^{-1}$ Raman spectral lines taken with the  NMR
	tube with the coating material and Cs vapor (Fig.~\ref{fig:tube}~II) after heating (a) in comparison with control NMR tube without metal (b)  (Fig.~\ref{fig:tube}~III). Chemical structures corresponding to peaks  shown on the top of figure. R -- hexadecyl group (--$\rm C_{16}H_{33}$) }
	\label{fig:cs-tube-approx}
\end{figure}

According to \cite{Socrates2004} the two higher-energy  peaks correspond to the trans- ($\rm 1673~cm^{-1}$) and cis- ($\rm 1659~cm^{-1}$ ) isomers of a non-terminal alkene. 
Figure~\ref{fig:cs-tube-approx} shows these three peaks for sample II (alkene with Cs, after heating) fit to the sum of three Gaussians, along with the spectrum of sample III (alkene without Cs) for comparison.
The ratio of their areas is 59/28/14 for trans-, cis- isomers and alpha-olefin for sample II (after heating), compared to 100\% alpha-olefin in sample III.

There were no changes observed in the tube sample without alkali metal, which was treated the same way as the sample with Cs, indicating that all changes in the spectrum were related to interactions of the coating material with the Cs vapor.

Because all spectra except the one for the coating on the wafer (Fig.~\ref{fig:r-spectra}e)  were taken through the glass of the cells or tubes, some of the spectral features near $\rm 1250~cm^{-1}$ and $\rm 500~cm^{-1}$  originate from glass. A spectrum of glass is shown in Fig.~\ref{fig:r-spectra}f.

\begin{figure*}
	\centering
		\includegraphics[width=0.9 \linewidth]{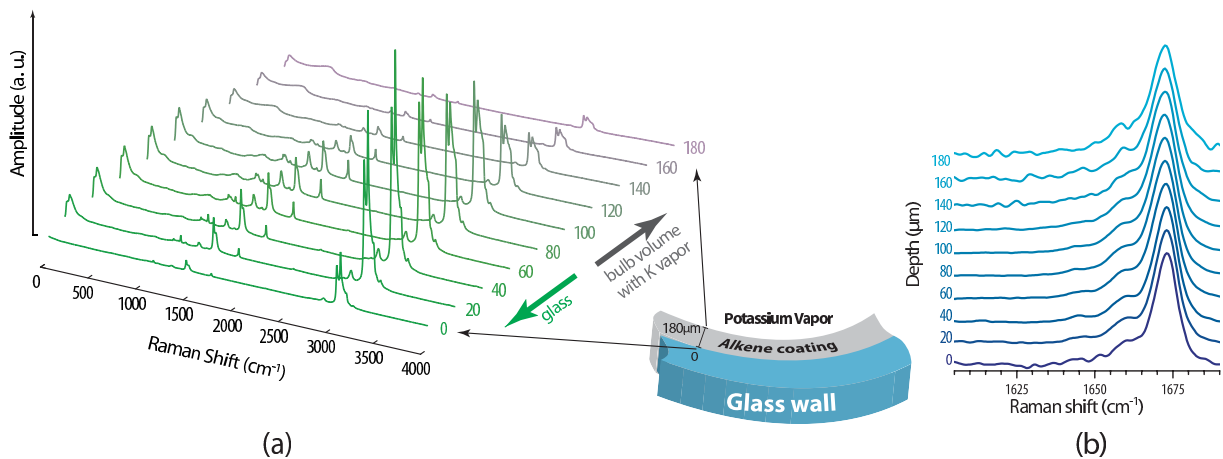}
	\caption{ 
	a) “Raman spectra as a function of displacement from the wall of a potassium-vapor alkene-coated cell (Sample I). All spectra recorded with the same exposure.
	b) Zoomed region 1605 - 1690  $ \rm cm^{-1}$. Each spectrum in (b) is normalized with respect to 1673 $ \rm cm^{-1}$ peak.
	}
	\label{fig:profiling}
\end{figure*}

The dependencies of signal amplitude, peak positions and other Raman-spectrum features on the depth of the focal plane position into the coating for potassium cell with alkene coating were also investigated. The results are presented in Fig.~\ref{fig:profiling}a. 
The spectroscope's confocal microscope was focused on the cell's glass. Spectra were measured 10 times moving the focus into the coating bulk in 20~$\rm \mu m$ steps.  The pump-light  wavelength was 532~nm and  the power was  20~mW. Exposures were the same for all spectra. The picture  on the right shows the zoom of the 1673 $\rm cm^{-1}$ region. 
For estimation of the dependence of the ratio of trans-, cis- isomers and alpha-olefin on the depth, spectral region 1605 - 1690  $ \rm cm^{-1}$ is zoomed on Fig.~\ref{fig:profiling}a. Each spectrum is normalized to the 1673 $\rm cm^{-1}$ peak.
The results show that the whole bulk of the coating experiences interaction with  alkali metal.

It is difficult to interpret the low-wave number (frequency) part of the spectrum. 
The $\rm 112~cm^{-1}$ line is absent in Fig.~\ref{fig:r-spectra}~a and present for the same sample on each of the spectra in Fig.~\ref{fig:profiling}. This may be due to the spectral  proximity of the pump light.

\section{Experimental results: NMR \label{NMR}} 

\begin{figure}
	\centering
		\includegraphics[width=1.0 \linewidth]{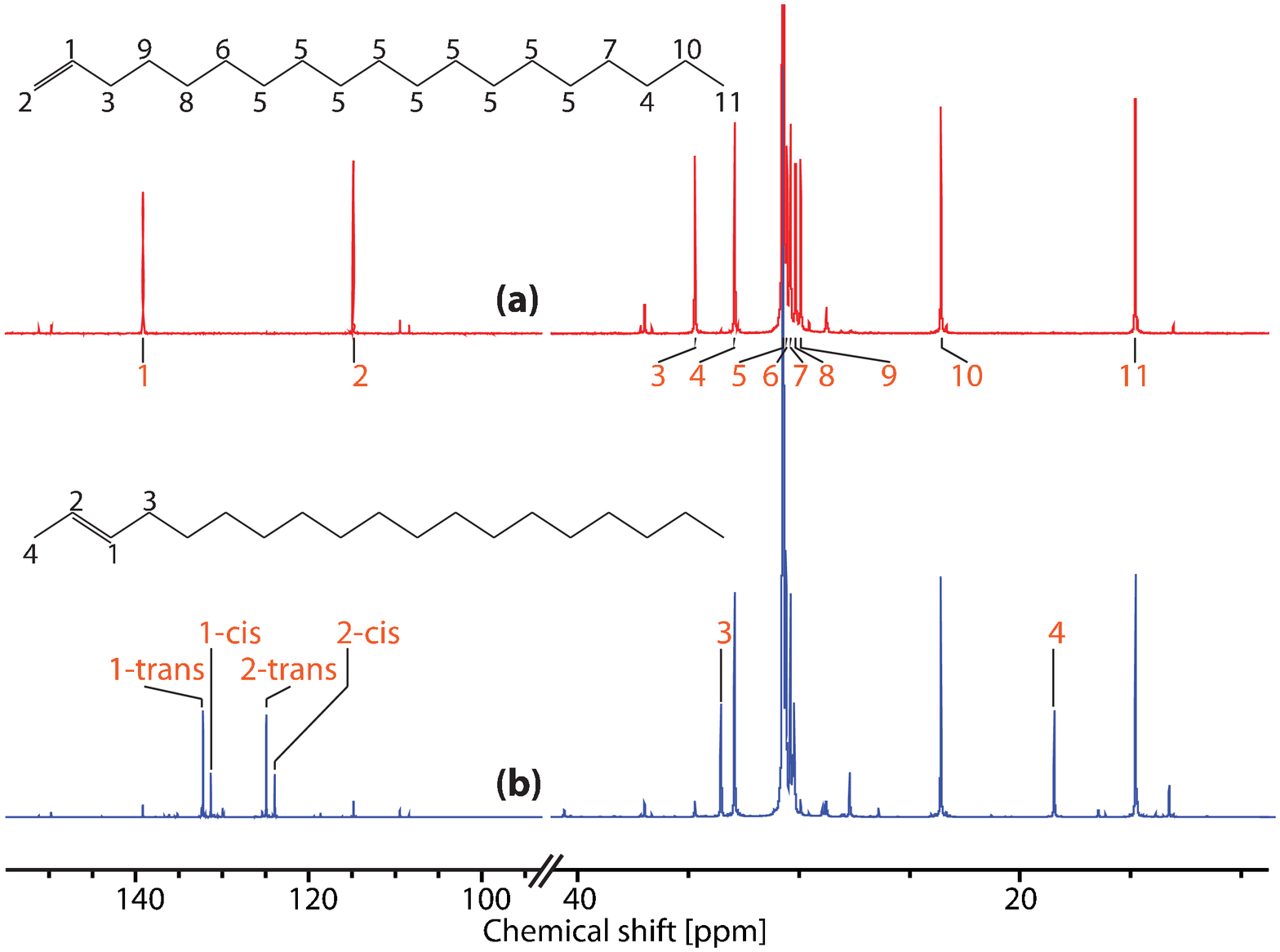}
	\caption{$^{13} \rm C$ NMR spectra: a -- NMR tube with the coating material and Cs vapor (Fig.~\ref{fig:tube}~II) immediately after preparing (the same spectrum for the control tube with the coating material only (Fig.~\ref{fig:tube}~III) is not presented), b -- NMR tube with the coating material and Cs vapor (Fig.~\ref{fig:tube}~II) after heating at $80 \rm ^\circ \rm C$ during one month. The numbers marking the spectral lines associate peaks to atom position in the molecule \cite{Mistry2009}. 
	}
	\label{fig:C-NMR}
\end{figure}

\begin{figure}
	\centering
		\includegraphics[width=1.0 \linewidth]{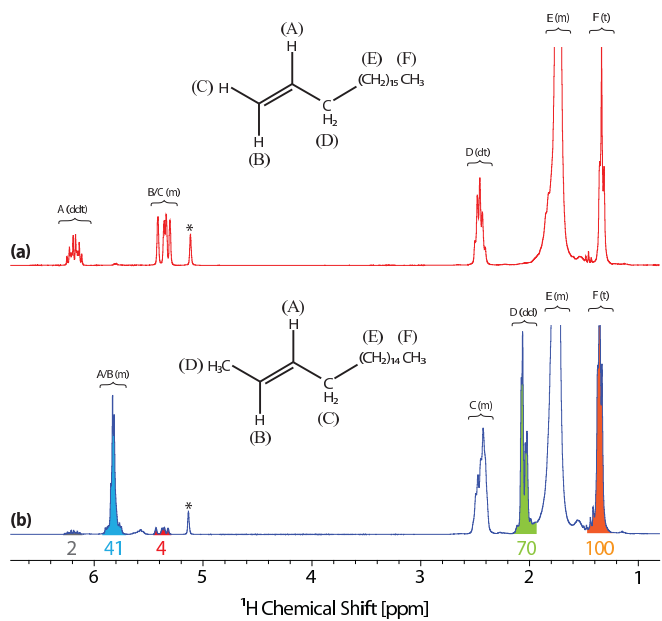}
	\caption{ $^{1}\rm H$ NMR spectra: a -- NMR tube with the coating material and Cs vapor (Fig.~\ref{fig:tube}~II) immediately after preparing (the same spectrum for the control tube with the coating material only (Fig.~\ref{fig:tube}~III) is not presented), b -- NMR tube with the coating material and Cs vapor (Fig.~\ref{fig:tube}~II) after heating at $80 \rm ^\circ \rm C$ during one month.  The letters above the spectrum associate the peak groups with atoms' positions in the molecule \cite{Mistry2009}. The numbers under the colored areas are the relative areas of the peaks groups.}
	\label{fig:H-NMR}
\end{figure}

The NMR spectra of samples II and III confirm the above Raman results regarding the  double-bond migration. The spectra for pure coating material (sample III) and coating material with cesium (sample II) immediately after preparation were identical for  both  $^{1} \rm H$ (Fig.~\ref{fig:H-NMR}a) and $^{13} \rm C$ (Fig.~\ref{fig:C-NMR}a)  nuclei. 
However, upon heating of sample II, clear differences arise.

On the $^{13} \rm C$ spectrum of the sample for the tube with Cs after heating (Fig.~\ref{fig:C-NMR}b), the peaks at $\rm115~ppm$ and $\rm 139~ppm$ decreased in size, and two new pairs of peaks corresponding to cis- ($ \rm 131~ppm$ and $\rm 124~ppm$)  and trans- ($ \rm 132~ppm$ and $\rm 125~ppm$) isomer appeared. The NMR peaks were fit with Lorentzians \cite{Guenther2013}, and the ratios of the integrated-peak intensities are 66/27/7 for trans-, cis- isomers and alpha-olefin, respectively. These ratio are similar to the Raman spectrum for this sample. From the $^{1}\rm H$ spectrum the area ratios of $\rm -CH_3$-group from different ends of the alkene molecule can be calculated. The green colored D peak group  on Fig.~\ref{fig:H-NMR}b corresponds to the $\rm -CH_3$ group near the carbon double bond. Such a group is present in all isomerized products in this study. The orange colored  F peak group on Fig.~\ref{fig:H-NMR}b is characteristic of  any alkene molecule. 
The areas under the F and D peak groups are in the ratio 7/10. At the same time, the ratio of the areas of A/B peaks of isomerized alkene to B/C of residual initial material is 4/41. This  means that count of isomerized molecules is 91\%, which is different from estimation above by F and D peak.

\section{DISCUSSION\label{DISCUSSION}} 

Analysis of the NMR and Raman spectra indicates that the process of cell curing involves an alkali-metal-catalyzed isomerization of 1-nonadecene to 2-nonadecene. Extracting precise quantitative concentrations and product ratios is challenging, but we estimate that approximately 90\% of the $\alpha$-olefin reacted, preferentially yielding trans- over cis-2-nonadecene in a roughly 2:1 ratio.

The observed isomerization may be considered as a [1,3]-sigmatropic hydride shift as described in Ref.~\cite{Woodward2013}. However, geometrical constraints on the transition state predicted by the Woodward-Hoffman rules suggest that [1,3]-sigmatropic hydride shifts are forbidden in the absence of a catalyst or a photo-activated diradical mechanism. Because samples II (1-nonadecene with Cs) and III (1-nonadecene control) were both subject to the same illumination, photochemical mechanisms can be ruled out, indicating that the reaction is catalyzed by the alkali metal.

It is worth noting that a similar process has been studied in Ref. \cite{Pines1955}. Metallic sodium or its compounds were used in that work, but the base-catalysts mechanism was proposed in \cite{Pines1960} for reactions promoted by sodium compounds only.

An exact mechanism for the alkali-catalyzed sigmatropic rearrangement has not been proposed in previous work. However, transition-metal-promoted alkene isomerization has been observed: an adsorption catalytic mechanism was proposed in  \cite{Anderson1984}, and a reaction pathway through intermediate compounds was presented in \cite{Pettit1969}.

The most likely reaction pathway in the present case is through a short lived alkali metal – alkene complex which can lift the restriction of the Woodward-Hoffman rule, perhaps in a promoter-free analogue of the mechanism suggested in Ref.~\cite{Pines1955}.

There is no indication of stable metal-organic complexes as suggested in Ref.~\cite{Balabas2010}.

\section{CONCLUSIONS\label{CONCLUSIONS}} 

We reported here the results of our experimental investigation of transformation of an anti-relaxation coatings due to its interaction with alkali atoms. The investigations were made by Raman spectroscopy and NMR. It was found that alkene molecules undergo isomerization as a result of the interaction with the alkali atoms. 
We are convinced that this process is not a thermal or photochemically activated [1,3]-sigmatropic hydrogen shift.  
It is a result of a catalytic alkene interaction with alkali-metal atoms. A presence of metal-organic compounds suggested in Ref.~\cite{Balabas2010} was not confirmed. 
We refrain from offering speculative explanations as to benefits of the observed $\alpha$- to $\beta$-olefin isomerization in terms of vapor cell anti-relaxation coating performance, but we do suggest that it may be important for future studies of anti-relaxation coatings to consider alkali-metal-catalyzed reactions such as the one presented here.

\section{ACKNOWLEDGEMENTS\label{ACKNOWLEDGEMENTS}} 

We wish to thank Prof. A. Vasilyev of the chemical faculty of Saint-Petersburg State University for directing us towards sigmatropic mechanism of isomerization  and very helpful discussion of chemical mechanisms of the described processes and interest in our work. Dmitriy Budker acknowledges support by the National Science Foundation under award CHE-1308381, by the NGA NURI program, and by the DFG DIP project Ref. FO 703/2-1 1 SCHM 1049/7-1. Presentation of the results at the International Conference on Atomic Physics 2014 in Washington~DC~(USA) was supported  by Russian Science Foundation (grant \#14-12-00094). J.W.B. was supported by a National Science Foundation Graduate Research Fellowship under Grant No. DGE-1106400.
\bibliography{bibliography}

\end{document}